\documentclass[aps,twocolumn,superscriptaddress]{revtex4}

\usepackage[dvips]{graphicx}
\usepackage[]{amsmath}
\usepackage[]{amssymb}
\usepackage[]{amsfonts}
\usepackage[]{latexsym}
\usepackage[]{float}

\newcommand{\mb}[1]{\mbox{\boldmath $#1$}}

\newcommand{\mats}[4]
{\scriptstyle
\left(
\begin{array}{cc}
#1 & #2 \\
#3 & #4 
\end{array}
\right)
}

\begin{document}

\title{
Quantized Berry Phases as Local Order Parameters of Quantum Liquids
}

\author{Yasuhiro Hatsugai}
\date{Feb.23, 2006}
\affiliation{Department of Applied Physics, University of Tokyo, 7-3-1 Hongo Bunkyo-ku, Tokyo 113-8656, Japan}

\begin{abstract}
We propose to use quantized Berry phases as local 
order parameters
of gapped quantum liquids,
 which are invariant under some anti-unitary
operation. After presenting a general prescription, 
the scheme is applied
for Heisenberg models with frustrations
and  two dimensional
extended  Su-Schrieffer-Heeger models
associated with a random dimer covering.
In each phases,
the quantized Berry phases, as $0$ or $\pi$,
describe the ground state texture pattern of  
 local singlet bonds and dimer bonds.
Also possible applications to large 
classes of correlated electron systems are 
discussed.
\end{abstract}

\maketitle

One of the challenges in modern  condensed matter physics is to
have well understanding of quantum liquids which do not have 
classical analogues.
States of matters in  classical systems are mostly described 
by  the local order parameters based on a concept of symmetry breaking.
However, recent studies in decades 
have revealed many of interesting quantum phenomena
are not well characterized by the classical local order parameters,
such as quantum Hall effects, %
 exotic superconductors\cite{Laughlin98-idxy} and 
frustrated or doped quantum magnets\cite{Anderson87,Sachdev-book,Read91-LN}. 
Quantum spins with $S=1/2$
 and fermions can be objects in  quantum limits
which do not have classical correspondents. 
When they get together, one may find some classical degree of
the freedom to describe the states approximately.
However, it is not always the case.
A pair of  $S=1/2$ spins with  exchange interaction
forms a singlet and a triplet.
The latter has a classical analogue as a small magnet, however,
 the singlet does not. 
When the total system is composed of such  singlet 
pairs, the system is also in the quantum limit as a singlet spin liquid.
The most famous singlet spin liquid can be the RVB state proposed by Anderson 
as a possible basic platform of the high-$T_C$\cite{Anderson87}.
Also the valence bond solid  (VBS) state and 
the  Haldane phases of
integer Heisenberg spin chains
 are the quantum spin liquids 
of this class\cite{Haldane83-c,Affleck87-AKLT,Read91-LN}.%
 The ground state of a half filled Kondo lattice  also belongs 
to it, 
which is a superposition of singlet pairs between the
conduction electrons and the localized spins\cite{Tsunetsugu92}. 
Some of the dimer models and spins with a string net condensation can be 
solvable limits of such quantum liquids\cite{Rokhasar88,Sondi01,WenString}.
As for a fermion pair, 
when the hybridization between them is assumed  secondary,
 one may use a classical 
number basis for the description.
 However, the picture  breaks down 
in the strong coupling limit, where the state is labeled as bonding or
anti-bonding states. 
They are purely quantum variables, as bonds, that live on the link between the fermions sites.
This state with the dimerized fermion pairs can be also understood as 
a typical quantum liquid.

In these quantum liquids, the classical local order parameter is not
enough to characterize the state.
We proposed to use  generalized topological numbers 
such as the Chern numbers
for the {\em global} characterization based on a concept of
 topological orders\cite{wen89,Hatsugai04e}.
Also non local objects such as entanglement entropies
 can help to capture some of the features\cite{ent}.
Here we propose to use Berry phases to
characterize the quantum liquids {\em locally}. 
The Berry phase is gauge 
dependent and is not quantized generically\cite{berry84}.
We clarify the gauge dependence and present a prescription to define {\em quantized }
Berry phases as  {\rm quantum local order parameters}
 for {\em anti-unitary invariant } states.
This requirement for the invariance can be complementary to the one for the Chern number description
where the time-reversal symmetric states are mostly classified by
vanishing topological integers.

{\em Berry phases with  Gauge Fixing:}
Let us first present a prescription to
 calculate Berry phases for 
a multiplet which is a generalization of
a single eigen state to an eigen  space
with  dimension $D\ge 1$.\cite{Hatsugai04e} 
The multiplet naturally appears in a discussion 
with degeneracy and is also quite useful
to handle a single many body state
of the Slater state. 
(See later).
Let us consider a parameter $x$-dependent hermite hamiltonian which is
 diagonalized by ortho-normalized eigen states with
energies $E_j(x)$
as 
$
\mb{H} (x) \mb{\Psi} (x) = \mb{\Psi} (x) \mb{E}(x)
$,
$
\mb{\Psi} (x) = (| \Psi_1(x) \rangle , \cdots, |\Psi_D (x) \rangle )
$,
$
\mb{\Psi} ^\dagger \mb{\Psi}  = \mb{I} _D
$ 
and
$
\{\mb{E}(x)\}_{ij} =  \delta_{ij}  E_{j}(x)
$.
We further assume the gap opening condition,
$
^\forall x,\ E_j (x)\neq E_k(x)
$, 
$
j = 1,\cdots, D
$,
$k = D+1,\cdots$,
which guarantees a regularity of the multiplet
as for $x$\cite{Hatsugai04e}.
A matrix valued Berry connection,
$
\mb{A}  = \mb{\Psi} ^\dagger  d \mb{\Psi}
$ and
a closed loop $C$ in the parameter space define
the Berry phase,
which is customarily written 
as 
$
i{\gamma } _C (\mb{A} )= \int_C {\rm Tr \,} \mb{A} 
$.\cite{berry84,Wilczek84}
Changing a basis for the multiplet,
the multiplet is transformed by a unitary matrix $\mb{\omega} $ as 
$\mb{\Psi}=\mb{\Psi}^\prime \mb{\mb{\omega} }  $.
It induces a gauge transformation
$\mb{A}=\mb{\omega} ^\dagger \mb{A}^\prime \mb{\omega}  + \mb{\omega} ^\dagger d \mb{\omega}  $.
\cite{berry84,Hatsugai04e,Wilczek84}.
The Berry phase, $\mb{\gamma }_C $,
is gauge dependent and
thus is not well defined without a specific gauge fixing.
Following a general procedure\cite{Hatsugai04e}, the gauge
can be fixed by a multiplet $\mb{\Phi} $,
which  is  arbitrary and single-valued but not necessarily constant.
We define an unnormalized multiplet,
$
\mb{\Psi} ^U = \mb{P} \mb{\Phi} 
$
  by
the gauge invariant projection
$
\mb{P}  = \mb{\Psi}  \mb{\Psi} ^\dagger
$.
It is only normalized as
$\mb{\Psi} _\Phi = \mb{\Psi}^U \mb{N} _\Phi^{-1/2}$
with the gauge independent requirement
$\det \mb{N}_\Phi \neq 0$
where
$
 \mb{N}_\Phi= (\mb{\Psi}^U) ^\dagger \mb{\Psi}^U
= \mb{\eta}_\Phi ^\dagger \mb{\eta}_\Phi
$,
$ 
\mb{\eta}_\Phi = \mb{\Psi} ^\dagger \mb{\Phi}  
$ 
and
$
\det \mb{N}_\Phi = |\det \mb{\eta}_\Phi|^2
$.
When it is satisfied everywhere
on a curve $C$,
we call the gauge by the $\mb{\Phi} $ is {\em regular}.
Taking the other regular gauge 
by the multiplet $\mb{\Phi}^\prime $,
the transformation matrix is explicitly given as
$ \mb{\omega } =
(\mb{\Psi}_{\Phi^\prime}) ^\dagger\mb{\Psi}_{\Phi}
=
   \mb{N}_{\Phi'}^{-1/2}  (\mb{\eta}_{\Phi'} )^\dagger  
\mb{\eta}_\Phi   
   \mb{N}_{\Phi}^{-1/2} 
$ 
which  is
also regular
on the loop $C$\cite{footnote1}. 
The Berry phase gets modified as 
$
{\gamma } _{C} (\mb{A}_\Phi )= 
{\gamma } _{C} (\mb{A}_{\Phi^\prime} )+
 \Delta (C,\mb{\omega})
$,
$
  \Delta (C, \mb{\omega})=
   \int_C 
  d \text{Arg}\det\nolimits_D \mb{\omega }
=
  \int_C 
  d \text{Arg}  \det\nolimits _D \mb{\eta }_{\Phi} /\det\nolimits_D \mb{\eta }_{\Phi'}
$.
Since the overlap matrices $\mb{\eta}_{\Phi,\Phi'} $
are
single-valued and regular on the $C$,
we have $\Delta(C,\mb{\omega} )=2\pi M_C   $ with some integer $M_C$.
(Generically we expect $M_C=1$, but one may have $M_C>1$ with
some additional symmetry.)
It implies that the Berry phase in a regular gauge
 ({\em regular Berry phase})
 has a gauge
 invariant meaning up to this integer as 
$
\gamma _C \equiv 
-i \int_C\, {\rm Tr \,} \mb{A}$,
($\text{mod}\, 2\pi M_C$).
This is topological since small perturbations can not modify the 
Berry phase unless the gauge becomes singular.

{\em Anti-Unitary Symmetry and Quantized Berry phases:}
Now let us consider 
 a parameter independent 
discrete symmetry described by an {\em anti-unitary} operator
 $\Theta=K U_\Theta$ where
$K$ is a complex conjugation and 
$U_\Theta$ is  unitary.
It operates for a state
expanded by a parameter independent basis,
$|G(x) \rangle  = \sum_J C_J(x) | J \rangle $,
 as 
$
| G^\Theta(x) \rangle =  \Theta |G(x) \rangle = \sum_J C_J^*(x) | J^\Theta \rangle
$, 
where 
$\{|J^\Theta \rangle = \Theta |J \rangle \}$ are assumed to  form an orthonomalized basis.
This simple observation brings an important restriction to the  Berry phase as 
$
\gamma _C (G) = -\gamma _C(G^\Theta)
$. %
Now we assume that the %
hamiltonian $H(x)$ is invariant under the 
anti-unitary operator 
 $
[H,\Theta] = 0
$.
When the eigen state $|G\rangle $ is {\em unique}, 
$|G \rangle $ and $|G^\Theta \rangle $
are 
different  just in their phases (gauges). 
 They span the same (one dimensional)
linear space. %
With the generic argument before,
it implies 
$
\gamma _C (G) \equiv  \gamma _C(G^\Theta)$,
$(\text{mod}\, 2\pi M_C)$.
To be compatible with
the transformation property with $\Theta$,
the Berry phase of the anti-unitary invariant state is 
{\em inevitably quantized}
as
$
\gamma _C(G) = 0, \pi M_C$,
$(\text{mod}\ 2\pi M_C)$.
As for an example, let us consider
a  hamiltonian
with anti-unitary symmetry, $\Theta_D= K \sigma _x $,
$
H^{D}(x) = 
\mats
{0}{x}
{x^*}{0}
$,
$\Theta_D ^{-1} \mb{H} ^D \Theta_D =  \mb{H} ^D$
and
$x=e^{i\varphi}$.
A closed curve to parametrized the hamiltonian is a unit circle 
$C=\{e^{i\varphi}|\varphi:0\to 2\pi\}$. 
It is diagonalized as 
$
\mb{H}  ^D = \mb{P} _+-\mb{P} _-
$
where
$
\mb{P} _\pm = \mb{\Psi}_ \pm  \mb{\Psi}_ \pm ^\dagger 
$
 and
$
^t\!\mb{\Psi}_  \pm = 
({e^{i\varphi} },{\pm 1})/{\sqrt{2} } 
$.
Taking a 
gauge specified by a constant 
 $
^t\!{\mb{\Phi}}  = 
({a},{b})
$,
the overlap determinant for the negative energy multiplet 
is evaluated as 
$\det\nolimits_1\mb{\eta}_\Phi 
=
(a e^{-i\varphi} -b)/\sqrt{2}
=
e^{-i\varphi} (a-be^{i\varphi} )/\sqrt{2}
$
which never vanishes unless $|a|=|b|$.
It implies the gauge is regular if $|a|\neq |b|$.
Now let us write the gauge by $\Phi$ for $|a|>|b|$ and
the one by $\Phi^\prime $ for $|a|<|b|$.
These two gauges can not be continuously connected 
and thus the transformation matrix,
 $\mb{\omega} $, defines a {\em large} gauge transformation.
The  normalized multiplet for each gauges is evaluated as 
$
^t\!\mb{\Psi} _\Phi =
e^{i\text{Arg}(a-b e^{i\varphi} )} ({1},{-e^{-i\varphi} })/\sqrt{2}
$ and
$
^t\!\mb{\Psi} _{\Phi^\prime} =
e^{i\text{Arg}(b-a e^{-i\varphi} )} ({-e^{i\varphi} },1)/\sqrt{2}
$ respectively.
Each of them defines the regular Berry phase
as $\gamma _C(\mb{A}_\Phi) = -\pi $ 
and
 $\gamma _C(\mb{A}_{\Phi^\prime}) = \pi $.

{\em Local Topological Order Parameters:}
We propose to use the quantized Berry phases 
to define a {\em local} topological order parameter for 
gapped quantum liquids.
\begin{figure}
\begin{center}
\includegraphics[width=0.6\linewidth,clip]{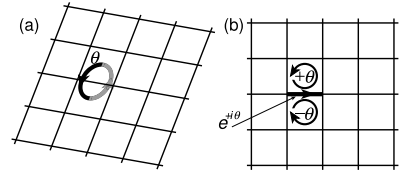}%
\end{center}
 \caption{
A flux loop as a local probe to define 
quantized Berry phases for picking up a local topological order.
(a) Schematic figure. (b) a local gauge to define the flux loop.
\label{f:local_ord}}
 \end{figure}
Let us consider a quantum many body state $| G \rangle  $ which is an
 eigen state of a hamiltonian $H$.
We do not require a translational symmetry and the system may have  boundaries as well.
We just assume an anti-unitary invariance of the hamiltonian by the  operator $\Theta$. 
Further we assume that the (ground) state is {\em gapped } and  {\em unique}, i.e.,
it is {\em invariant under the anti-unitary operation}.
Then introducing a {\em local} perturbation 
at $r$ 
specified by a set of parameters $x_{r}$,
let us define a hamiltonian $H(x_r)$ which preserves the symmetry requirement as 
$[H(x_r),\Theta] = 0$,
$\Theta| G(x_r) \rangle  \propto | G(x_r) \rangle $.
(Fig. \ref{f:local_ord}).
The local perturbation can be any. However, possible choices are some type of gauge 
potentials which can be gauged away locally, ( of course, not globally). 
Taking some closed curve $C_r$ in the parameter space,
we have two possibilities
:(a) $|G \rangle $ is 
unique ${^\forall x_r\in C_r}$ or (b)
$|G \rangle $ is  degenerate at $^\exists x_r\in C_r$.
In the first case, we identify 
the quantized Berry phase $\gamma ({C_r})$ as 
the local order parameter at $r$.
When the degeneracy occurs as in the latter case, 
 it implies 
 gapless localized excitations ({\em edge states})
are induced by the perturbation\cite{Hatsugai93b,ryu02}.
(In a fermion bilinear hamiltonian system, 
it is an appearance of localized zero modes. 
)
This also characterizes the location $r$ of the
quantum state $| G \rangle $.
Then we assign  one of the {\em three } labels,
$\{ 0$, $\pi M_C$,   ``gapless'' $\}$ to the position $r$.
We propose to use them as  local topological order parameters
for the gapped quantum liquid.

To evaluate  the quantized Berry phase 
for some specific models numerically,
 one needs to specify the gauge and 
discretize the loop $C$ 
as 
$
x_0,x_1\cdots x_N =x_0
$.
Then we define a lattice Berry phase $\gamma_C^N(\mb{A} ^\Phi ) $ 
by the lattice Berry connection\cite{Hatsugai04e,Fukui05}
as
 $
\gamma _C^N(\mb{A} ^\Phi)
= \text{Arg} \prod_{j=1}^N\, \det\nolimits_D
{ \mb{\Psi}_j^U} ^\dagger \mb{\Psi} _{j+1} ^U
$,
$
\mb{\Psi}^U _j = \mb{P}_j \mb{\Phi} 
$,
$
\mb{P}_j =  \mb{\Psi} _j \mb{\Psi} _j ^\dagger 
$
and 
$
\mb{\Psi} _j = \mb{\Psi} (x_j)
$.
In the large $N$ limit, it reduced to the standard one.
A generic choice of the
multiplet $\mb{\Phi} $ is expected to induce
 a regular gauge. In the numerical calculations below,
 we take random vectors for them.

\begin{figure}
\begin{center}
\includegraphics[width=0.42\linewidth,clip]{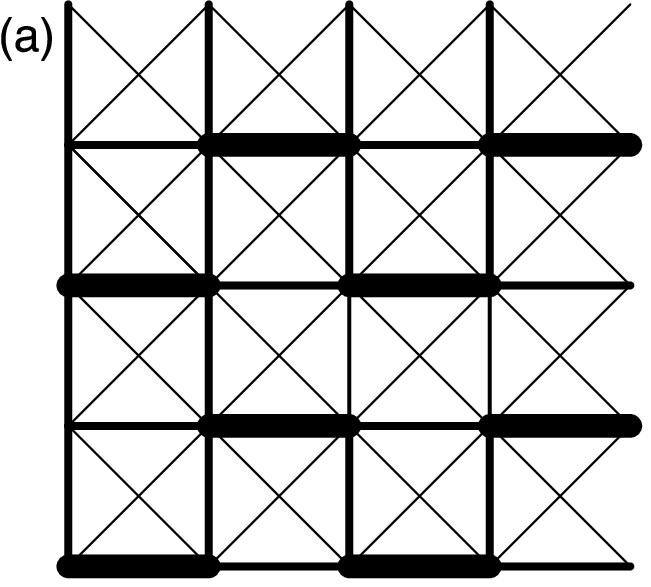}%
\includegraphics[width=0.42\linewidth,clip]{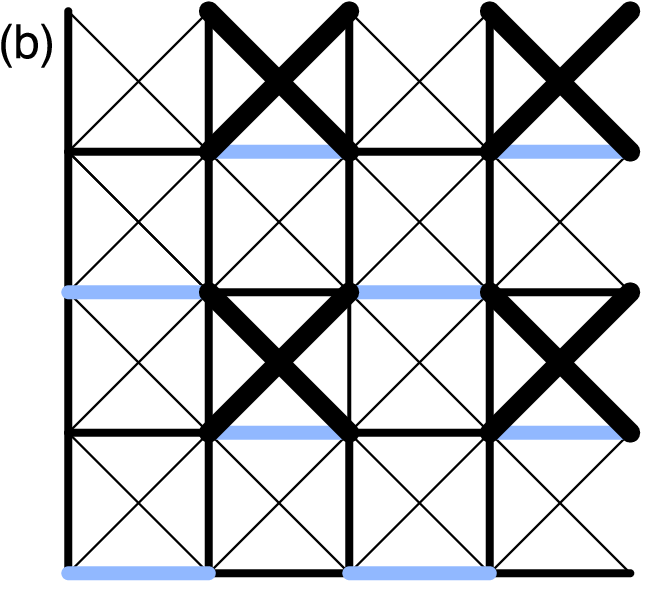}%
\end{center}
 \caption{
 A distribution of $\pi$-bonds
for  frustrated Heisenberg models on a  $4\times 4$ lattice with periodic boundary condition.
The most thick lines denote the $\pi$-bonds.
The other links are $0$-bonds. 
(a) 
Next-Nearest-Neighbor exchanges are
 $J_{NNN}=0.5$,
nearest neighbor exchanges are $J^S_{NN}=2$ or $J^W_{NN}=1$. The links with 
$J^S_{NN}=2$ coincide with the $\pi$-bonds
 denoted by the thick lines.
(b) 
Some of the $J_{NNN}$'s are enhanced from 0.5 to 4.
The other are the same as (a).
The strong NNN links $J^S_{NNN}=4$ coincide
with the $\pi$-links. Dimerized links, $J^S_{NN}=2$, ( the same as (a)) are 
shown as gray lines,
which are 0-bonds in this phase. (They are $\pi$-bonds in (a).)
\label{f:spin}
}
\end{figure}

{\em Heisenberg Models on a Frustrated Lattice:}
We apply the generic formulation here for specific models.
The first example is  a generic Heisenberg model on any lattice in any dimensions. 
We allow frustrations among spins.
The hamiltonian  is given as 
$
H^{sp} =  \sum_{ij}
{J }_{ij}{\mb{S} _i} \cdot
\mb{S} _j 
$
where $^t\!{\mb{S}_i} =(S_{ix},S_{iy},S_{iz})$, $S_{i \alpha }=\frac {1}{2}  {\sigma }_{i \alpha }$,
$\alpha =x,y,z  $ and
$\mb{\sigma} $'s are the Pauli matrices.
A time-reversal operator,
 $\Theta_T=K \otimes_j(i \sigma _{jy})$,  is used to describe 
the anti-unitary symmetry.
It operates for a generic spin state 
$
|G \rangle = \sum C_J | \sigma _1,\sigma _2,\cdots,\sigma _N \rangle
$,
($\sigma _i=\pm 1$)
as 
$
\Theta_{ T} |G \rangle = \sum C_J^* (-)^{\sum_{i=1}^N (1+\sigma _i)/2}
| -\sigma _1,\cdots,-\sigma _N \rangle 
$.
The spins get transformed as
$
^\forall j,\ \mb{S}_j \to  \Theta_{ T} ^{-1}  \mb{S}_j \Theta_{ T} = - \mb{S} _j
$.
It is thus clear the Hamiltonian  is invariant as 
$
\Theta_{ T} ^{-1} 
H^{sp}
\Theta_{ T} 
=  H^{sp}
$.
A local perturbation to define the quantized Berry phase is constructed 
by a local gauge transformation
$
\mb{S}_j\to {\mb{S}}_j'(\varphi_j) \equiv  e^{i \varphi_j \mb{n}\cdot \mb{S }_j  }
 \mb{S} _j
e^{-i \varphi \mb{n}\cdot \mb{S }_j  } 
= \mb{Q}_j (\varphi)\mb{S} _j
$
where $\mb{Q}(\varphi)=e^{\varphi\mb{T}  }$ is a real $3\times 3$  matrix and
the  $\mb{T} $  is  real skew symmetric as 
$
T^{\alpha \beta } = 4i n ^ \gamma {\rm Tr \,} S^\alpha S^\beta S^\gamma 
$\cite{Hatsugai04e}.
We take $\hat {\mb{n} }=(0,0,1)$ without losing a generality.
Then the local hamiltonian is written in the transformed basis as 
 $ J_{ij}\mb{S} _i \cdot \mb{S} _j
=h_{ij}(\mb{S}_i',\mb{S}_j',\theta_{ij}  )=
 \frac {1}{2} 
(
e^{- \theta_{ij} }
S_{i+} 'S_{j-}'
+
e^{ \theta_{ij}}
S_{i-}'S_{j+}'
)
+  S_{iz}'S_{jz}'
$
where $\theta_{ij}=
\varphi_i
-
\varphi_j
$.
Based on the observation,
let us consider a Heisenberg model
with a local perturbation  at the link $\langle i_0j_0\rangle $ as 
$
H^{sp}(x_{\langle i_0j_0 \rangle }=e^{i\theta})
=\sum_{ij}h_{ij}(\mb{S}_i,\mb{S}_j,\theta_{ij}  )
$,
where
${\theta} _{ij} =\theta$ for $\langle ij \rangle =\langle i_0j_0  \rangle $ and $0$ otherwise.
Then consider a quantized Berry phase 
$\gamma _C^{sp}$ by the unit circle $C$.
Note that this local twist is not gauged away globally and induces a local perturbation
which preserves the time-reversal symmetry.
Let us first consider a special configuration of the interactions $J_{ij}$.
Taking any nearest-neighbor dimer covering of all sites 
 ${\cal D}=\{\langle ij \rangle  \}$
($\#{\cal D}=N/2$, $N$ is a number of total sites),
we assume that the interaction is  nonzero only on these dimer links as 
$
H^{sp}= \sum_{\langle ij \rangle  \in {\cal D}} J_{ij} \mb{S}_i\cdot \mb{S}  _j
$ ($^\forall J_{ij}>0$).
The ground state $|G \rangle $ is unique and gapped. 
It is also invariant as for the time-reversal operation.
When  the link $\langle i_0j _0\rangle  $ does not belong to 
the dimer covering ${\cal D}$, the Berry phase is apparently vanishes
 $\gamma _{C_{  \langle i_0 j_0 \rangle   }}^{sp}=0$.
When 
 the dimer covering includes the link $\langle i_0j _0\rangle  $,
the  ground state in the regular gauge is explicitly given as 
$
|G _\Phi \rangle 
=\big[
|i_0j _0\rangle
 \mb{\Psi}  _\Phi(e^{i\theta}) 
\big]
\bigotimes_{\langle ij \rangle \neq \langle i_0j _0\rangle   } 
\big[
|{ij}\rangle  \mb{\Psi}  _\Phi(1)
\big] 
$
where
$
|{ij}\rangle = 
(
|+_i \rangle |-_j  \rangle ,
|-_i \rangle |+_j  \rangle )
$.
Then the quantized Berry phase is evaluated as $\pi$ similar to the example ($H^D$).
It is clear that the quantized Berry phases pick up the dimer pattern ${\cal D}$ as
$
\gamma _{C_{\langle ij \rangle }} = \pi: \langle ij \rangle \in {\cal D}$ and 
$
\gamma _{C_{\langle ij \rangle }}= 0: \langle ij \rangle \notin {\cal D}$.
Now let us imagine an adiabatic process to include interactions across
 the dimers.
Due to the topological stability of the quantized Berry phase,
they can not be modified unless the dimer gap collapses.
This  dimer limit presents 
a non-trivial example and shows the usefulness
of the quantized Berry phases as {\em local order parameters 
of singlet pairs}. 
To show its real validity of the quantized Berry phases,
we have diagonalized the Heisenberg hamiltonians numerically by
the Lanzcos algorithm and calculated $\gamma _{C_{\langle ij \rangle }}$
numerically.
Some of the  results are shown in Figs.\ref{f:spin} as 
a demonstration.
Clearly the
quantized Berry phases are quite powerful to describe  texture patterns 
of the singlet liquid phases\cite{Hatsugai06-detail}.

{\em Dimerization of Spinless fermions and the Slater States:}
Another example of the anti-unitary operator is a 
particle-hole symmetry operation for fermion systems.
To be specific let us discuss a spinless fermion with a particle-particle 
interaction by
a hamiltonian 
$
  H^{sl} (x) = \sum_{\langle ij \rangle } 
 t_{ij}(x)c_i ^\dagger c_j +
 t_{ij}^*(x)c_j ^\dagger c_i
 +
 V_{ij}(x) (n_i-\frac {1}{2} )( n_j-\frac {1}{2} )
$,
where  $c_i$ is a fermion annihilation operator.
We divide $N$ lattice sites into $A$ and $B$ sublattices.
The particle-hole conjugation operator,  $\Theta_{ P}$,
which is anti-unitary 
 is defined by
$\Theta_{ P} = K U_{ P} $,
$U_{ P}  =
 \overrightarrow{\prod}_j\xi_{j}
$,
$
 \xi_j =  
{\xi_{j+}}(^\forall j\in A),
\
{\xi_{j-}}
(^\forall j\in B)$
where $\xi_{j+}=\xi_{j+} ^\dagger =(c_j+c_j ^\dagger )$ and
$\xi_{j-}=\xi_{j-}^\dagger =-i(c_j-c_j ^\dagger )$ 
are Majorana fermions,
 $\{\xi_\alpha ,\xi_\beta \}=2 \delta_{\alpha \beta } $,
$
\xi_{j\pm} \ c_j\ \xi_{j\pm } = \pm c_j ^\dagger 
$.
When the lattice is {\em bipartite} and 
 the hopping $t_{ij}$ connects sites between the sublattices $A$ and $B$, 
the hamiltonian is anti-unitary invariant by $\Theta_P$ as 
$
\Theta_{ P} ^{-1} \, H^{sl}(x)\, \Theta_{ P} = H^{sl}(x)
$.
Thus a generic  $M$-particle  state, $|G_N \rangle $  is degenerate with
the $(N-M)$-particle state,
$
|G_{N-M} \rangle
=\Theta_P|G_N \rangle 
$.
It implies that the  eigen space of the $\Theta_P$ operator
is not one-dimensional except the half filled case ($N=2M$).
In this half filled case, 
we apply the present characterization by the quantized Berry phase 
assuming the state is unique and gapped.
We take a local perturbation at a link $\langle i_0j_0\rangle $
similarly to the Heisenberg spins
as
$
t_{ij }= |t_{ij}| e^{i\theta_{ij}}$,
where
${\theta} _{ij} =\theta$ for $\langle ij \rangle =\langle i_0j_0  \rangle $ and $0$ otherwise.
Taking a unit circle  $C$,
the quantized Berry phase is evaluated.
We have performed exact diagonalizations similar to the Heisenberg spins
\cite{Hatsugai06-detail}.
However, we show here further analysis assuming the system is non-interacting $V_{ij}=0$.
When we assume that the hopping is only non zero among
the  dimers in some dimer covering ${\cal D}$, the quantized Berry phases 
reproduce the dimer pattern.
This model
 can be understood 
as a generalization of the  Su-Schrieffer-Heeger (SSH) model 
in one-dimension\cite{Heeger88,ryu02}.
In the dimer limit, 
the particle-hole symmetric unique ground  state is 
given as
$
|G _\Phi \rangle 
=\big[
 \mb{c} _{\langle i_0j _0\rangle  } ^\dagger \mb{\Psi}  _\Phi(e^{i\theta}) 
\big]
\prod_{\langle ij \rangle \neq \langle i_0j _0\rangle   } 
\big[
\mb{c}_{ij} ^\dagger \mb{\Psi}  _\Phi(1)
\big] | 0 \rangle 
$
where
$
 \mb{c} _{ij} ^\dagger = (c_i ^\dagger ,c_j ^\dagger )
$.
The regular Berry phase is then evaluated as
$
\gamma _{C_{\langle ij \rangle}  }(G_\Phi)
= 
{\pi}:$ 
$\langle ij \rangle  \in {\cal D}$,
$0:$ (otherwise).
Due to the topological stability, this texture pattern is only modified
after a quantum phase transition.
The quantized Berry phases here play a role of 
 {\em topological order parameters } of the {\em local dimerization}.

\begin{figure}
\begin{center}
\includegraphics[width=0.49\linewidth,clip]{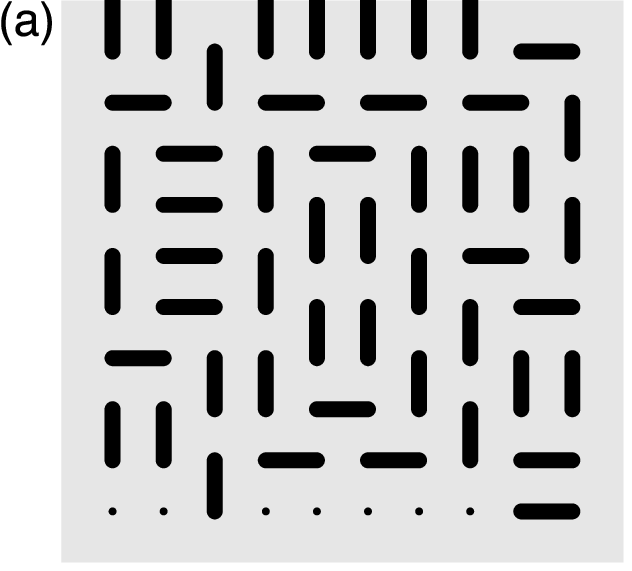}
\includegraphics[width=0.49\linewidth,clip]{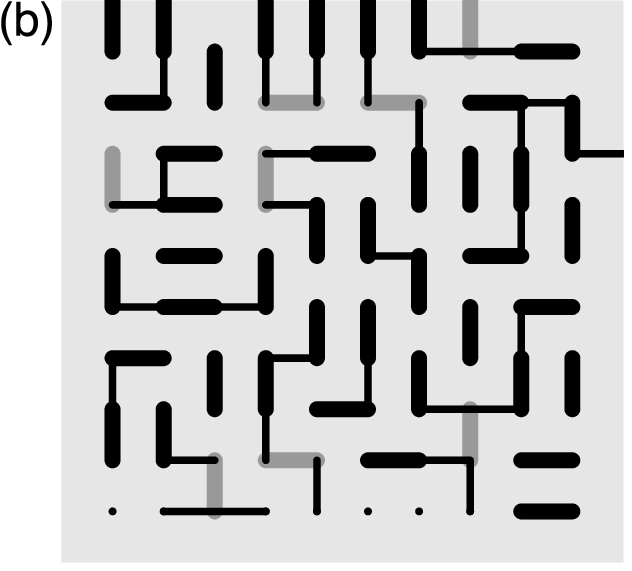}
\end{center}
 \caption{
Non interacting spinless fermions with hopping modulations,
$|t_{ij}|=t_S$ or $t_W\, (\le t_S)$.
The strong links $t_S$ are distributed on a (random)
dimer covering configuration.
The black thick lines are the strong links 
 with $\gamma =\pi$.
The gray thick lines are the strong links 
with $\gamma =0$.
The black thin lines are the weak links 
 with $\gamma =\pi$.
The other nearest neighbor links, which are not drawn,
are weak links with $\gamma =0$.
(a) $t_W=0.6t_S$
and (b) $t_W=0.7t_S$. The dimer configuration is the same. 
There occur several quantum phase transitions
supplemented with  gap closings
 between $t_W:0.6\rightarrow 0.7$.
\label{f:rnd}}
 \end{figure}

In this non-interacting case, the  non-Abelian formulation here
is quite useful for the evaluation of the Berry phase of 
a {\em single many body state}.
The $M$-particle eigen state (Slater state) is constructed from  a set of the 
one-particle states
as  (a generic fermi sea)
$
| G_M \rangle  = 
|\{ \mb{\varphi}\} \rangle =
 \prod_{\ell=1}^M ( \mb{c} ^\dagger \mb{\varphi} _\ell )| 0 \rangle
$,
$
\mb{c}^\dagger  = (c_1 ^\dagger ,\cdots,c_N ^\dagger )
$,
where $|0 \rangle $ is a vacuum $^\forall i,c_i| 0 \rangle =0$.
The ortho-normalized 
one-particle state $\mb{\varphi}_\ell $, 
$(
\mb{\varphi}_\ell ^\dagger 
\mb{\varphi}_{\ell ^\prime} = \delta_{\ell\ell^\prime} 
) $ is 
an eigen state of the one-particle hamiltonian $\mb{h} $
where 
$\mb{h}\mb{\varphi}_\ell = \epsilon_{\ell} \mb{\varphi}_\ell  $ and 
$\mb{c} ^\dagger   \mb{h} \mb{c}  = {H} ^{sl}$, ($V_{ij}= 0)$.
As for this $M$-particle state, 
a multiplet with a dimension $D=M$, 
$\mb{\varphi} =(\mb{\varphi}_1,\cdots,\mb{\varphi}_M)  $ 
defines a non-Abelian Berry connection 
$
\mb{A}_M=\mb{\varphi} ^\dagger  d  \mb{\varphi}
 $, 
which gives the Berry connection
 of  the single many-body state $| G_M \rangle $
as 
$A=  \langle G_M| d G_M \rangle 
 = {\rm Tr \,} \mb{A} _M
$\cite{Hatsugai04e,Hatsugai06-detail,footnote2}.
It enables us to perform  calculations of the 
Berry phases for quite large systems
where the non-Abelian gauge fixing is crucial.
We have demonstrated its validity for an extended SSH model in two dimensions
on a  $10\times 10$ periodic lattice as shown in  Fig.\ref{f:rnd}.
The $\pi$-bonds
 pick up the dimer pattern of the model in the dimmer limit and 
also captured topological structure of the complicated quantum liquid 
after several quantum phase transitions.

{\em The $tJU$  model and the Kondo lattice:}
The present generic scheme can be useful for large classes of correlated 
electron systems.
For example,
following $tJUV$ models with Kondo 
coupling
in $d$ dimensions are worth to be investigated by the scheme,
$
H^{tJUVK}(x) = \sum_{\langle ij \rangle } (H^T_{ij} + H^V_{ij} + H^J_{ij} )
+\sum_i( H^U_i +H^K_i)
$,
$
H^T_{ij} (x)= 
 \mb{c}_{i } ^\dagger  \mb{t}_{ij}(x)  \mb{c} _{j}  + h.c. 
$,
$
H^V_{ij} = (n_i-1) V_{ij}(n_j-1)
$,
$H^J_{ij} (x)= \, 
 ^t\!\mb{s}  _i
 \mb{J }_{ij}^J (x)
\mb{s}  _j
$,
$
H_i^U = U (n_{i\uparrow} -1/2)(n_{i\downarrow} -1/2)
= -U\frac {2}{3} \mb{s}_i ^2
$,
$
H^K(x) = ^t\!\mb{s}_i \mb{J}_i^K(x)   \mb{S} _i
$ and 
$
\mb{J }_{ij}^{J,K} (x)={J_{ ij  }^{J,K}}^0\mb{Q} (\theta^{J,K}_{ij} )
\text{diag}(1,1,\lambda^{J,K}_{ij })
$,
where $\mb{c}_i ^\dagger  = 
(
c_{i \uparrow } ^\dagger,
c_{i \downarrow } ^\dagger
 ) $, 
$\mb{s}_i = \frac {1}{2}  \mb{c}_i ^\dagger \mb{\sigma } \mb{c}_i   $ 
and $^t\!{\mb{s}_i}=({s}_{ix},{s}_{iy},{s}_{iz} )  $.
In the model, two kinds of anti-unitary operators,
the particle-hole conjugation $\Theta_P$ and the
time-reversal operation  $\Theta_T$  can be  the symmetric operations.
They are explicitly defined as 
$
\Theta_P = K U_P
$,
$
U_P = U_P ^+ U_P ^-
$,
$
U_P ^\sigma =\overrightarrow \prod_j
\xi_{j \sigma } 
$
and 
$
\Theta_T =K \prod_i
e^{\frac {\pi}{2}(
s_{i+}
-
s_{i-}
)}
$.
By these anti-unitary operations, we can consider
two different type of local perturbations as discussed.
Then the quantized Berry phases are defined as
two different local topological order parameters.
They can be useful to describe and characterize quantum liquids phases
of the correlated electron systems.
\cite{Hatsugai06-detail}
 The ground state of 
the half filled Kondo lattice  is expected to give a 
non trivial example.
At least, in the strong coupling limit,
the Kondo coupling links can be labeled as $\pi$-bonds
 as for the time-reversal 
symmetry. 

Part of the  present work  was supported by Grant-in-Aid from
Japanese Ministry of Science and Culture and the Sumitomo Foundation.


\vskip -0.1cm

\begin{thebibliography}{21}

\expandafter\ifx\csname natexlab\endcsname\relax\def\natexlab#1{#1}\fi
\expandafter\ifx\csname bibnamefont\endcsname\relax
  \def\bibnamefont#1{#1}\fi
\expandafter\ifx\csname bibfnamefont\endcsname\relax
  \def\bibfnamefont#1{#1}\fi
\expandafter\ifx\csname citenamefont\endcsname\relax
  \def\citenamefont#1{#1}\fi
\expandafter\ifx\csname url\endcsname\relax
  \def\url#1{\texttt{#1}}\fi
\expandafter\ifx\csname urlprefix\endcsname\relax\def\urlprefix{URL }\fi
\providecommand{\bibinfo}[2]{#2}
\providecommand{\eprint}[2][]{\url{#2}}

\bibitem[{\citenamefont{Laughlin}(1998)}]{Laughlin98-idxy}
\bibinfo{author}{\bibfnamefont{R.~B.} \bibnamefont{Laughlin}},
  \bibinfo{journal}{Phys.\ Rev.\ Lett.} \textbf{\bibinfo{volume}{80}},
  \bibinfo{pages}{5188} (\bibinfo{year}{1998}).

\bibitem[{\citenamefont{Anderson}(1987)}]{Anderson87}
\bibinfo{author}{\bibfnamefont{P.~W.} \bibnamefont{Anderson}},
  \bibinfo{journal}{Science} \textbf{\bibinfo{volume}{235}},
  \bibinfo{pages}{1196} (\bibinfo{year}{1987}).

\bibitem[{\citenamefont{Sachdev}(1999)}]{Sachdev-book}
\bibinfo{author}{\bibfnamefont{S.}~\bibnamefont{Sachdev}},
  \emph{\bibinfo{title}{Quantum Phase Transition}}
  (\bibinfo{publisher}{Cambridge Univ. Press}, \bibinfo{year}{1999}).

\bibitem[{\citenamefont{Read and Sachdev}(1991)}]{Read91-LN}
\bibinfo{author}{\bibfnamefont{N.}~\bibnamefont{Read}} \bibnamefont{and}
  \bibinfo{author}{\bibfnamefont{S.}~\bibnamefont{Sachdev}},
  \bibinfo{journal}{Phys.\ Rev.\ Lett.} 
 \textbf{\bibinfo{volume}{62}},
  \bibinfo{pages}{1694} (\bibinfo{year}{1989}),
\textbf{\bibinfo{volume}{66}},
  \bibinfo{pages}{1773} (\bibinfo{year}{1991}).

\bibitem[{\citenamefont{Haldane}(1983)}]{Haldane83-c}
\bibinfo{author}{\bibfnamefont{F.~D.~M.} \bibnamefont{Haldane}},
  \bibinfo{journal}{Phys.\ Lett.} \textbf{\bibinfo{volume}{A93}},
  \bibinfo{pages}{464} (\bibinfo{year}{1983}).

\bibitem[{\citenamefont{Affleck et~al.}(1987)\citenamefont{Affleck, Kennedy,
  Lieb, and Tasaki}}]{Affleck87-AKLT}
\bibinfo{author}{\bibfnamefont{I.}~\bibnamefont{Affleck}},
  \bibinfo{author}{\bibfnamefont{T.}~\bibnamefont{Kennedy}},
  \bibinfo{author}{\bibfnamefont{E.~H.} \bibnamefont{Lieb}}, \bibnamefont{and}
  \bibinfo{author}{\bibfnamefont{H.}~\bibnamefont{Tasaki}},
  \bibinfo{journal}{Phys.\ Rev.\ Lett.} \textbf{\bibinfo{volume}{59}},
  \bibinfo{pages}{799} (\bibinfo{year}{1987}).

\bibitem[{\citenamefont{Tsunetsugu et~al.}(1992)\citenamefont{Tsunetsugu,
  Hatsugai, Ueda, and Sigrist}}]{Tsunetsugu92}
\bibinfo{author}{\bibfnamefont{H.}~\bibnamefont{Tsunetsugu}},
  \bibinfo{author}{\bibfnamefont{Y.}~\bibnamefont{Hatsugai}},
  \bibinfo{author}{\bibfnamefont{K.}~\bibnamefont{Ueda}}, \bibnamefont{and}
  \bibinfo{author}{\bibfnamefont{M.}~\bibnamefont{Sigrist}},
  \bibinfo{journal}{Phys.\ Rev.\ B} \textbf{\bibinfo{volume}{46}},
  \bibinfo{pages}{3175} (\bibinfo{year}{1992}).

\bibitem[{\citenamefont{Rokhsar and Kivelson}(1988)}]{Rokhasar88}
\bibinfo{author}{\bibfnamefont{D.~S.} \bibnamefont{Rokhsar}} \bibnamefont{and}
  \bibinfo{author}{\bibfnamefont{S.~A.} \bibnamefont{Kivelson}},
  \bibinfo{journal}{Phys.\ Rev.\ Lett.} \textbf{\bibinfo{volume}{61}},
  \bibinfo{pages}{2376} (\bibinfo{year}{1988}).

\bibitem[{\citenamefont{Moessner and Sondhi}(1989)}]{Sondi01}
\bibinfo{author}{\bibfnamefont{R.}~\bibnamefont{Moessner}} \bibnamefont{and}
  \bibinfo{author}{\bibfnamefont{S.~L.} \bibnamefont{Sondhi}},
  \bibinfo{journal}{Phys.\ Rev.\ Lett.} \textbf{\bibinfo{volume}{86}},
  \bibinfo{pages}{1881} (\bibinfo{year}{1989}).

\bibitem[{\citenamefont{Wen}(2003)}]{WenString}
\bibinfo{author}{\bibfnamefont{X.~G.} \bibnamefont{Wen}},
  \bibinfo{journal}{Phys. \ Rev.\ Lett.} \textbf{\bibinfo{volume}{90}},
  \bibinfo{pages}{016803} (\bibinfo{year}{2003}).


\bibitem[{\citenamefont{Wen}(1989)}]{wen89}
\bibinfo{author}{\bibfnamefont{X.~G.} \bibnamefont{Wen}},
  \bibinfo{journal}{Phys. \ Rev.\ B} \textbf{\bibinfo{volume}{40}},
  \bibinfo{pages}{7387} (\bibinfo{year}{1989}).



\bibitem[{\citenamefont{Hatsugai}(2004)}]{Hatsugai04e}
\bibinfo{author}{\bibfnamefont{Y.}~\bibnamefont{Hatsugai}},
  \bibinfo{journal}{J.\ Phys. \ Soc.\ Jpn.} \textbf{\bibinfo{volume}{73}},
  \bibinfo{pages}{2604} (\bibinfo{year}{2004}),
 \textbf{\bibinfo{volume}{74}},
  \bibinfo{pages}{1374} (\bibinfo{year}{2005}).

\bibitem{ent}
S. Ryu and Y. Hatsugai, cond-mat/0601237.

\bibitem[{\citenamefont{Berry}(1984)}]{berry84}
\bibinfo{author}{\bibfnamefont{M.~V.} \bibnamefont{Berry}},
  \bibinfo{journal}{Proc.\ R.\ Soc.} \textbf{\bibinfo{volume}{A392}},
  \bibinfo{pages}{45} (\bibinfo{year}{1984}).


\bibitem{Wilczek84}
F. Wilczek and A. Zee,
Phys. \ Rev.\ Lett. 
\textbf{\bibinfo{volume}{52}}, 2111 (1984). 








\bibitem{footnote1}
$
\mb{\omega} ^\dagger \mb{\omega} 
=
   \mb{N}_{\Phi}^{-1/2}  (\mb{\eta}_{\Phi} )^\dagger  
\mb{\eta}_{\Phi'}
   \mb{N}_{\Phi'}^{-1} 
  (\mb{\eta}_{\Phi'} )^\dagger  
\mb{\eta}_\Phi   
   \mb{N}_{\Phi}^{-1/2} 
=\mb{I} _D
$

\bibitem{Hatsugai93b}
Y. Hatsugai,
Phys. \ Rev.\ Lett. 
\textbf{\bibinfo{volume}{71}}, 3697 (1993),
\bibitem{ryu02}
S. Ryu and Y. Hatsugai,
Phys. Rev. Lett. 
\textbf{\bibinfo{volume}{89}}, 077002 (2002).

\bibitem[{\citenamefont{Fukui et~al.}(2005)\citenamefont{Fukui, Hatsugai, and
  Suzuki}}]{Fukui05}
\bibinfo{author}{\bibfnamefont{T.}~\bibnamefont{Fukui}},
  \bibinfo{author}{\bibfnamefont{Y.}~\bibnamefont{Hatsugai}}, \bibnamefont{and}
  \bibinfo{author}{\bibfnamefont{H.}~\bibnamefont{Suzuki}},
  \bibinfo{journal}{J.\ Phys. \ Soc.\ Jpn.} \textbf{\bibinfo{volume}{74}},
  \bibinfo{pages}{1674} (\bibinfo{year}{2005}).


\bibitem[{\citenamefont{Hatsugai}(to be published)}]{Hatsugai06-detail}
\bibinfo{author}{\bibfnamefont{Y.}~\bibnamefont{Hatsugai}}
 (\bibinfo{year}{to be published}).


\bibitem[{\citenamefont{Heeger et~al.}(1988)\citenamefont{Heeger, Kivelson,
  Schrieffer, and Su}}]{Heeger88}
\bibinfo{author}{\bibfnamefont{A.}~\bibnamefont{Heeger}},
  \bibinfo{author}{\bibfnamefont{S.}~\bibnamefont{Kivelson}},
  \bibinfo{author}{\bibfnamefont{J.~R.} \bibnamefont{Schrieffer}},
  \bibnamefont{and} \bibinfo{author}{\bibfnamefont{W.~P.} \bibnamefont{Su}},
  \bibinfo{journal}{Rev.\ Mod.\ Phys.} \textbf{\bibinfo{volume}{60}},
  \bibinfo{pages}{781} (\bibinfo{year}{1988}).

\bibitem{footnote2}
$
 \langle G_M| d G_M\rangle 
= 
\langle \{ \mb{\varphi}\} | 
\sum_\ell
\cdots
(\mb{c}^\dagger \mb{\varphi}_{\ell-1}  )
(\mb{c}^\dagger d\mb{\varphi}_{\ell}  )
(\mb{c}^\dagger \mb{\varphi}_{\ell+1}  )
\cdots|0 \rangle 
= 
\sum_\ell \det\nolimits_M \mb{\varphi} ^\dagger \mb{} 
(\mb{\varphi}_{1},
\cdots
\mb{\varphi}_{\ell-1},
d\mb{\varphi}_{\ell},
\mb{\varphi}_{\ell+1}, \cdots )
=
\sum\nolimits_\ell \mb{\varphi} _\ell ^\dagger d\mb{\varphi} _\ell
={\rm Tr \,} \mb{\varphi} ^\dagger d \mb{\varphi}   $

\end{thebibliography}
\end{document}